\newcommand{\btem}{\bibitem}
\documentclass[12pt]{article}

\usepackage[dvips]{graphicx,color}
\begin{document}
\begin{center}
\begin{large}
{\bf The sigma meson and chiral restoration 
in nuclear medium}\end{large}
\\
\vspace{.5cm}
Teiji KUNIHIRO\\
\vspace{.5cm}
Yukawa Institute for Theoretical Physics, Kyoto University,
 Kyoto, 606-8502 Japan
\end{center}
\begin{small}
After giving  brief discussions on the 
 chiral condensate in hot and/or dense hadronic matter
 and the significance of the $\sigma$ meson in QCD,
 we discuss the importance of  experiments using nuclear targets
 including ones with electro-magnetic probes
 for obtaining  clearer confirmation of the existence of the
 $\sigma$ meson and for exploring the possible restoration of chiral
 symmetry in nuclear medium.
\end{small}

\section{Introduction}
The whole content of the present talk is based on the 
simple fact that chiral transition is  a phase transition of 
the QCD vacuum with the quark condensate 
$\langle \bar{q}q\rangle$ being the order parameter.
The relevance of the present talk to this
 workshop on physics with electro-magnetic probes 
is ensured by the fact that chiral
symmetry may be  at least partially restored 
in hot and/or dense hadronic medium, hopefully including
 finite nuclei.
In fact, a simple argument shows that the magnitude of the 
condensate $\langle\langle \bar{q}q\rangle\rangle_{T, \rho}$ at 
finite temperature $T$ and/or density $\rho$ tends to decrease.

To get convinced of this, one may first recall
 that the  Feynman-Hellman (F-H) theorem\cite{1} tells us that
the scalar charge of a hadron h $\langle \bar{q}_iq_i\rangle_h$
can be evaluated as a derivative of the hadron mass
$m_h$ with respect to the current quark mass $m_i$;
$\langle \bar{q}_iq_i\rangle_h= \partial m_h/\partial m_i.$
This is because the mass $m_h$ is an eigenvalue of
 the QCD Hamiltonian $H_{\rm QCD}$, which 
in turn contains the
current quark masses in the form $\sum_im_i\bar{q}_iq_i$.
As an almost trivial extension\cite{2,3,4}, the quark condensate 
in hot and dense
hadronic matter is given as a derivative of the free energy 
${\cal F}(T, \rho)$ of the system w.r.t. the current quark mass;
\begin{eqnarray}
\langle\langle\bar{q_i}q_i\rangle\rangle_{T,\rho}= \partial {\cal F}/\partial m_i
 &=&\partial {E_v}/\partial m_i+\partial {F}/\partial m_i\nonumber, \\
   & =&\langle\bar{q_i}q_i\rangle+
     \delta\langle\langle\bar{q_i}q_i\rangle\rangle_{T,\rho},
\end{eqnarray}
where we have assumed that 
${\cal F}(T, \rho)$ is given as a sum of
 the vacuum energy $E_v$ and
 the usual free energy $F$ of the hadron system
( without
 the vacuum energy); ${\cal F}=E_v+F$.

If $F$ is the free energy of a hot pion gas, for example, 
one readily obtains 
\begin{eqnarray}
\delta \langle\langle \bar{q_i}q_i\rangle\rangle_T
=\sum_pn_{\pi}(p)\langle\pi(p)\vert\bar{q}_iq_i\vert\pi(p)\rangle,
\end{eqnarray}
where
$n_{\pi}(p)$ is the Bose-Eistein distribution function
for the pion and 
$\langle\pi(p)\vert\bar{q}_iq_i\vert\pi(p)\rangle
=\partial E_p^{\pi}/\partial 
m_i=m_{\pi}\langle \bar{q}_iq_i\rangle_{\pi}/E^{\pi}_p$,
with $E_p^{\pi}=\sqrt{m_{\pi}^2+p^2}$ is the scalar
charge of the pion with momentum $p$.
Applying F-H theorem to Gell-Mann-Oakes-Renner relation, 
$ f_{\pi}^2m_{\pi}^2=-(m_u+m_d)/2\cdot\langle \bar{u}u+\bar{d}d\rangle$,
one finds that 
$\langle \bar{q}_iq_i\rangle_{\pi}=6.25>0$
($i=u, d$).
Thus,
the magnitude of $\langle\langle\bar{q_i}q_i\rangle\rangle_T$
decreases  at finite temperature;
hence a partial
 restoration of chiral symmetry occurs at finite temperature.

In much the same way, if $F$ is the free energy of 
 a degenerate cold nuclear matter, one has\cite{3}, 
\begin{eqnarray}
\langle\langle \bar{q}_iq_i\rangle\rangle _{\rho}
=\big(1 -\frac{\Sigma _{\pi N}}{m_{\pi}^2f_{\pi}^2}\rho
\big) \langle\bar{q}_iq_i\rangle,
\end{eqnarray}
where
$\Sigma _{\pi N}=(m_u+m_d)/2\cdot\langle N\vert \bar{u}u+\bar {d} d\vert 
N\rangle$
$\sim (40 - 50)$ MeV is $\pi$-N sigma term.
Thus, 
the quark condensate tends to decreases in the nuclear medium:
At the normal nuclear density $\rho_0=0.17$fm$^{-3}$,
we have about 30 \% reduction of the quark condensate.
\footnote{
Notice, however, that the higher order terms in the density 
expansion 
originating from  the interaction energy between nucleons
modify the leading order result.}

If a phase transition is of 2nd order or {\em weak} 1st order,
there exist soft modes, which 
are    the fluctuations of the order parameter\cite{5}.
 For the chiral transition,
the fluctuation of the order parameter 
$\langle(\bar{q}q)^2\rangle$ 
is a scalar-isoscalar meson, which is
 historically called the $\sigma$-meson.
Thus one sees that the $\sigma$ meson becomes
 the soft mode of chiral transition at $T\not=0$ and/or
$\rho_B\not=0$.\cite{6}
In this report, 
 we discuss the importance of 
nuclear experiments with nuclear targets
 including ones with electro-magnetic probes
 for obtaining  clearer confirmation of the existence of the
 $\sigma$ meson and for exploring the possible restoration of chiral
 symmetry in nuclear medium. But, what is the $\sigma$? 
%

\section{The significance of the $\sigma$ meson in QCD}

The significance of the $\sigma$
 meson in hadron and nuclear physics may be summarized as 
follows\cite{7}:\begin{enumerate}
\item The $\sigma$ meson is the quantum fluctuation of the order
parameter $\langle (\bar{q}q)^2\rangle$, hence analogous to 
{the Higgs particle in the standard model}:
 NJL-like models\cite{8,6} and
 mended symmetry of Weinberg\cite{9} predict that 
 the mass $m_{\sigma}=400$-800 MeV and the
width $\Gamma\sim m_{\sigma}$. \, 
\item Recent various analyses\cite{10} have revealed the existence
of a pole in the 
$\pi$-$\pi$ $S$-matrix in the $\sigma$ channel.
In these analyses,
significance of  respecting {\em chiral symmetry,
unitarity and crossing symmetry} to reproduce the phase shifts
 both in the $\sigma\, (s)$- and $\rho\, (t)$-channels
{with a low mass $\sigma$ pole} 
has been recognized\cite{11}.\,
\item Such a scalar meson has been known to be
 responsible for the intermediate range attraction in
the nuclear force\cite{12}. 
\,
\item There are attempts\cite{13} to show that such an collectiveness 
in the sigma channel can accounts for $\Delta I=1/2$ enhancement in
the decay  
K$^0\rightarrow 2\pi$ 
compared with K$^{+}\rightarrow \pi^{+}\pi^{-}$.\, 
\item The empirical value of $\pi$-N sigma term 
$\Sigma_{\pi N}\sim$ 40-50 MeV
 implies an enhanced scalar charge of the nucleon
$\langle \bar{u}u+\bar{d}d\rangle_N\sim 8-10$
provided that $(m_u+m_d)/2\sim 5.5$MeV,
 in comparison with the naive value 
$\langle \bar{u}u+\bar{d}d\rangle_N=3$.
The collectiveness as summarized
 as the existence of the $\sigma$ meson
 can account for such an enhancement of
 the scalar charge\cite{14}.

\item 
 It should be noticed that the difficulties of the linear $\sigma$
model\cite{15} does not necessarily deny the linear realization
 of chiral symmetry where the $\sigma$ meson 
appears\cite{16}.
\end{enumerate}

\section{Production of the $\sigma$-meson in nuclear medium}
Is the pole observed in the $\pi$-$\pi$ phase shift really 
the $\sigma$ as the quantum fluctuation of the
order parameter of the chiral transition? 
The answer could be obtained clearly if
 the temperature and/or density are freely changed and 
 variations of the 
excitation modes, i.e., hadrons, on top of the varied
 vacua are traced as is usually done in 
condensed matter physics.
The life is, unfortunately, not so easy with the
 QCD vacuum. 
Nevertheless one may notice that 
nuclei provide us with the baryon density and 
 expect that 
nuclei might be dense enough
 to give rise to a partial restoration of chiral symmetry!
Some years ago,  the present author\cite{7} 
proposed several nuclear experiments including
 one using electro-magnetic probes 
to try to produce the $\sigma$ meson in nuclei
to see a clearer evidence of 
the existence of the $\sigma $ meson and to explore possible
 restoration of chiral symmetry in nuclear medium.

What are good observables to see the softening in the 
sigma channel in nuclear medium?
When a hadrons is put in a nucleus,
the hadron may dissociate into complicated
excitation to loose its identity in the  medium;
for example, $\sigma\leftrightarrow 2\pi$, \quad
$\sigma\leftrightarrow $ p-h, $\pi+$p-h, $\Delta$-h, $\pi+\Delta$-h $...$
Then the most informative quantity is the response function 
or spectral function of the system.
Serious calculations  of the strength function were performed
 by Chiku and Hatsuda\cite{17} and Volkov et al\cite{18}:
Their finding is 
 an enhancement of the spectral
 function in the sigma channel near the 2$m_{\pi}$ threshold.
The surprise was
such an enhancement had been seen by an experiment by 
CHAOS collaboration not at $T\not=0$ but at $\rho_B\not=0$
 in the cross sections of
A($\pi^{+}, \pi^{+}\pi^{\pm}$)A$'$\,
 (A$=2\rightarrow 208$)\cite{19};
 this experiment was actually motivated 
to explore possible in-medium $\pi$-$\pi$ correlations\cite{20}.

\hspace{-.2cm}
\begin{minipage}{8.5cm}
\scalebox{0.5}{%
\includegraphics{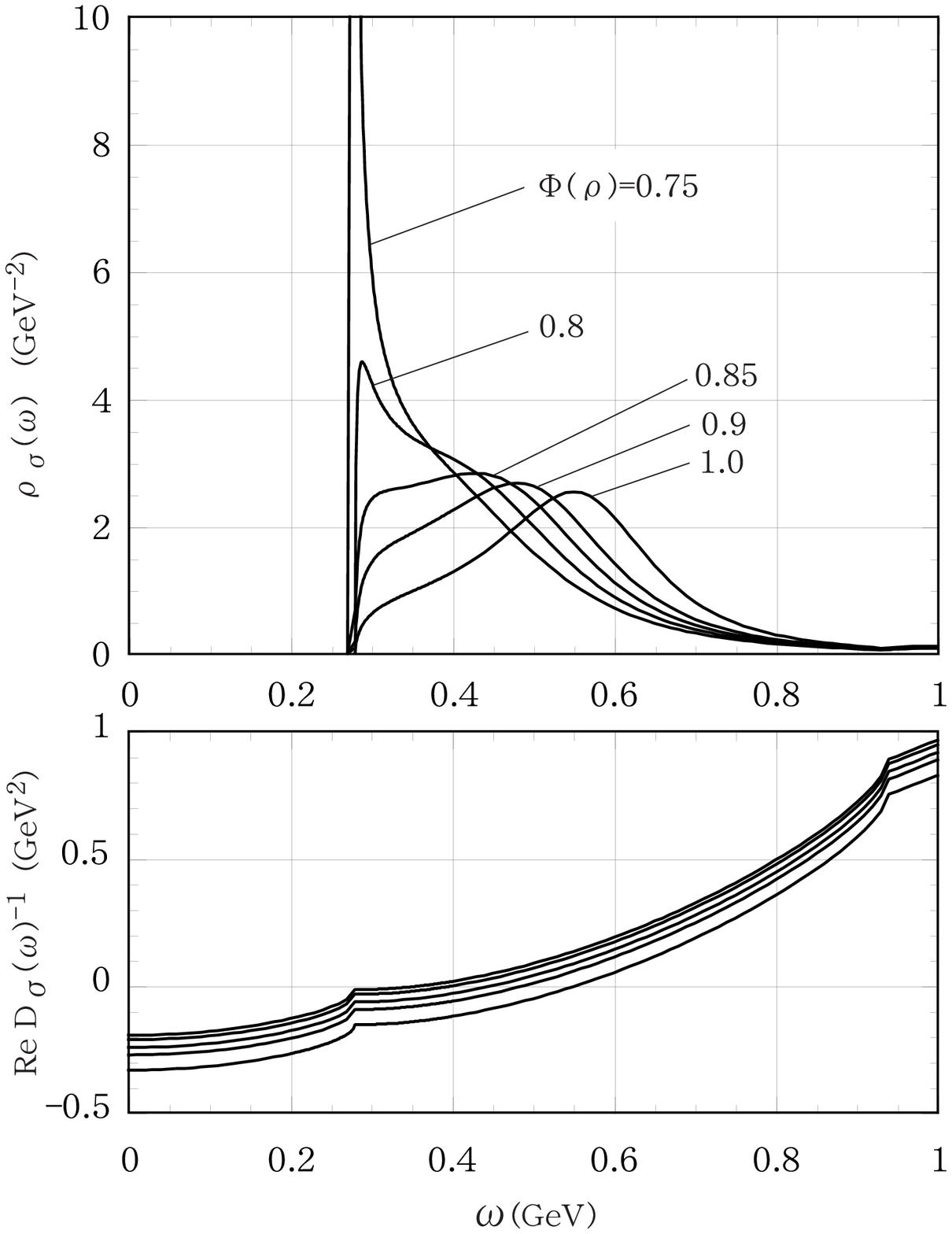}
}
Fig.1:\begin{small} The spectral function in the $\sigma$ 
channel\\
 (upper panel) and the inverse of
the $\sigma$ propagator.\\
 $\Phi$ denotes the 
ratio of the chiral condensates at $\rho$\\
 and in the vacuum.
\end{small}
\hfill
\end{minipage}
\begin{minipage}{7.5cm}
Then a calculation of the strength function at  $\rho_B\not=0$
 was made  using a linear sigma model\cite{21}.
It was shown that chiral restoration in the nuclear medium 
can lead to the required enhancement near the 
$2m_{\pi}$ threshold.
The result is shown in Fig.1: 
The upper panel shows the spectral functions in the $\sigma$ channel in 
nuclear medium with various densities parameterized by  $\Phi(\rho)$ 
denoting the ratio of the quark condensate at the density $\rho$ to 
the vacuum value.
 One can see the near-$2m_{\pi}$ threshold 
enhancement of the spectral function as the chiral symmetry is restored.
The lower panel shows the inverse of the Green's function of the
$\sigma$ propagator with the same $\Phi$ values; the upper curves corresponding
 to the smaller $\Phi$. 
One can see the distance between the cusps located at the
 $2m_{\pi}$-threshold and the zero decreases as the chiral symmetry is 
 getting restored, which causes the spectral enhancement near the threshold.
This was later confirmed by other groups\cite{22}.
\end{minipage}
One should notice that the
 state of the art calculation in the conventional
 reaction\break   
theory using nonlinear 
chiral Lagrangian but with no chiral restoration
 incorporated fail in
 reproducing the sufficient enhancement\cite{23}.

\section{ Discussion: The spectral enhancement in the
nonlinear chiral models }

How can   the near-threshold enhancement be described
 by non-linear chiral Lagrangians which are used in the
 conventional approaches?
An answer to
 this question has been given by Jido et al\cite{24}.
They showed the following:
(a) Although there is no explicit $\sigma$-degrees of freedom,
 there arises a decrease of the
 pion decay constant $f_{\pi}^{\ast}$ in nuclear medium.
(b) This is due to a new vertex, i.e.,
4$\pi$N-N
 vertex absent in the free space and in the
 previous calculations of the $(\pi, 2\pi)$ reaction; see Fig.2.
The vertex is responsible
for the reduction of  $f_{\pi}^{\ast}$ and hence for 
the spectral enhancement.
\begin{figure}[hbt]
\begin{center}
\scalebox{1.}{%
\includegraphics{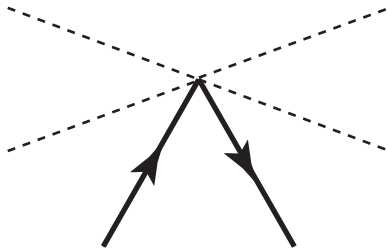}
}
\caption{
Fig.2: The new vertex causing
 the softening of the
spectral
 function
 in the $\sigma$ channel
 even in the 
 non-linear realization.
}
\end{center}
\end{figure}

They started with a linear sigma model as given below;
\begin{eqnarray}
\label{model-l}
{\cal L}  =   {1 \over 4} {\rm Tr} [\partial M \partial M^{\dagger}
 - \mu^2 M M^{\dagger} - {2 \lambda \over 4! } (M M^{\dagger})^2
 + h (M+M^{\dagger}) ]
  + \bar{\psi} ( i \partial \hspace{-6pt}/ - g M_5 ) \psi
  + \cdot \cdot \cdot  ,
\end{eqnarray}
where  $M = \sigma + i \vec{\tau}\cdot \vec{\pi}$,
 $M_5 = \sigma + i \gamma_5 \vec{\tau}\cdot \vec{\pi}$,
 $\psi$ is the nucleon field, and
 Tr is for the flavor index.
Making a polar decomposition 
$M = \sigma + i\vec{\tau}\cdot \vec{\pi}=
(\langle \sigma\rangle+S)U$ with 
$U={\rm e}^{i\vec{\tau}\cdot \vec{\pi}}$, 
(\ref{model-l}) is converted to 
 a nonlinear chiral Lagrangian with the massive scalar
field $S$;
\begin{eqnarray}
\label{model-nl}
{\cal L} & = &  {1 \over 2} [(\partial S)^2 - m_{\sigma}^{*2} S^2]
  - {\lambda \langle \sigma \rangle \over 6} S^3 - 
{\lambda \over 4!} S^4 
 +  {(\langle \sigma \rangle +S)^2 \over 4} {\rm Tr}
 [\partial U \partial U^{\dagger}] + 
{ \langle \sigma \rangle + S \over 4}\  h \
 {\rm Tr}[U^{\dagger}+U] \nonumber \\
& + & {\cal L}_{\pi N}^{(1)} - g S \bar{N} N \ ,
\end{eqnarray}
with
${\cal L}_{\pi N}^{(1)} =
\bar{N}(i \partial \hspace{-6pt}/ + i v \hspace{-6pt}/
 + i  a \hspace{-6pt}/   \gamma_{5}  - m_{N}^* ) N$,
where $(v_{\mu},a_{\mu}) = (\xi \partial_{\mu} \xi^{\dagger}
 \pm \xi^{\dagger} \partial_\mu \xi)/2$, and
 $m_N^* = g \langle \sigma \rangle$.

In the heavy-$S$ limit with 
 $g/\lambda$ and $\langle \sigma \rangle_0 = f_{\pi}$ fixed,
the heavy scalar field  $S$ may be integrated out
to give the following effective Lagrangian:
\begin{eqnarray}
\label{model-nl2}
{\cal L} & = &
 \left(
 {f_{\pi}^2 \over 4}
 - {g f_{\pi}  \over 2 m_{\sigma}^2}\bar{N}N
 \right)
\left(
{\rm Tr} [\partial U \partial U^{\dagger}]
 - {h \over f_{\pi}} \  {\rm Tr}[U^{\dagger}+U] \right).
\end{eqnarray}
The second term in the coefficient may be replaced
 by 
${g f_{\pi}  \over 2 m_{\sigma}^2}\rho,$
in nuclear medium, hence gives rise to a renormalization of 
$f_{\pi}$, which is renormalized away by  a redefinition of the 
pion field in the medium. Jido et al \cite{24} showed that 
this redefinition of the pion field in turn enhances
 the attraction between the pions in the $I=J=0$ 
channel leading to the softening of the spectral
 function in the $\sigma$ channel.

The  new vertex,
\begin{eqnarray}
\label{new-vertex}
{\cal L}_{\rm new} = - {3g \over 2 \lambda f_{\pi}} \
\bar{N}N {\rm Tr} [\partial U \partial U^{\dagger}],
\end{eqnarray}
depicted in Fig.2
has not been considered so far  in the calculations
of  the $\pi\pi$ scattering amplitudes in nuclear matter in the
 non-linear chiral Lagrangian approaches\cite{23}.


\section{Summary and concluding remarks}

The present talk may be summarized as follows:\\
\begin{enumerate}
\item The $\sigma$ meson as 
{\em the quantum fluctuation of the
 order parameter} of the chiral transition
 may account for various  phenomena in hadron physics
which otherwise remain mysterious.
\item There have been accumulation of 
{\em experimental
evidence of the
$\sigma$ pole} 
in the pi-pi scattering matrix. 
To deduce this result, it is found essential
 to respect chiral symmetry, analyticity and crossing symmetry.
\item Partial restoration of chiral symmetry in hot and dense medium
 leads to {\em a peculiar 
enhancement in the spectral function
 in the $\sigma$ channel near the $2m_{\pi}$ threshold}.
 \item Such an enhancement has been observed 
 in the reaction
A($\pi^{+}$, $(\pi^{+}\pi^{-})_{I=J=0}$)A' 
by CHAOS collaboration, which  may be 
{\em an experimental
evidence of the partial restoration 
of chiral symmetry in heavy nuclei}.
\item  The  spectral softening in the
 $\sigma$ channel is obtained both in the
linear and nonlinear realization of chiral symmetry
{\em 
provided that the
 possible reduction of the quark condensate or $f_{\pi}$
 is taken into account.}
\end{enumerate}
 
Further  works are necessary 
 to 
confirm that the near $2m_{\pi}$-threshold enhancement observed in 
the ($\pi^{+}$, $\pi^{+}\pi^{-}$) reactions by CHAOS collaboration
 is surely due to a partial restoration of chiral symmetry in nuclear
 medium:\, 
\begin{enumerate}
\item The strength function in the $\sigma$ channel in the wider 
$(\omega, q)$ region should be measured with various nuclear and 
electro-magnetic
 probes; 
for instance, photo- or electro-production of the $\sigma$ as well
 as the production by (d, $^{3}$He) and (d, t) reactions are interesting.
We remark that formation of $\sigma$ mesic nuclei\cite{25}
by (d, $^{3}$He) and (d, t) reactions are proposed as was done to
produce the deeply-bound pionic atoms\cite{25}.\,
\item To identify the $\sigma$ meson and the spectral function in 
 that
 channel, detecting 2$\pi^{0}$ and lepton pairs with $q\not=0$ are
 interesting. 
\end{enumerate}

\vspace{1pc}
This talk is largely based on the works
 done in collaboration with T. Hatsuda, D. Jido and H. Shimizu,
 to whom the author is very much grateful.

\end{document}